\documentclass[12pt]{article}
\usepackage{epsfig, axodraw}
\usepackage [small] {caption}
\newcommand{\ra}{\rightarrow}
\newcommand{\bq}{\begin{eqnarray}}
\newcommand{\eq}{\end{eqnarray}}
\newcommand{\p}{{\bf k}_{\perp}^2}
\newcommand{\ov}{\overline}

\begin{document}
\begin{center} {\bf Has the E791 experiment measured \\ the pion
    wave function profile ?}
\end{center}
\vspace{0.5cm}
\begin{center}{\bf Victor Chernyak}
\end{center}
\begin{center}
  Budker Institute of Nuclear Physics,\\
  630090 Novosibirsk, Russia
\end{center}
\vspace{1cm}
\begin{center}
  Abstract
\end{center}

The cross section of hard diffractive dissociation of the pion into
two jets is calculated. It is obtained that the distribution 
of longitudinal momenta for jets is not simply proportional to the 
profile of the pion wave function, but depends on it in a complicated way.  
In particular, it is shown that, under the conditions of the E791
experiment, the momentum distribution of jets is similar in its shape for
the asymptotic and CZ wave functions, and even the ratio of the
differential cross sections is not far from unity.

We argue therefore that, unfortunately, the E791 experiment has not
yet measured the profile of the pion wave function. For this, the
experimental accuracy has to be increased essentially.  
\vspace{1cm}

{\bf 1.} \hspace{1cm} The E791 experiment at Fermilab \cite{Aitala}
has recently measured  the cross section of the hard diffractive
dissociation of the pion into two jets. In particular, the
distribution of the total pion longitudinal momentum into fractions $y_1$ and
$y_2,\, (y_1+y_2)=1,\,$ between jets has been measured. The main purpose 
was to obtain in this way the information about the leading twist pion wave
function $\phi_{\pi}(x_1,x_2) $, which describes the distribition of
quarks inside the pion in the longitudinal momentum fractions 
$x_1$ and $x_2,\,(x_1+x_2)=1$.
  
The hope was based on the theoretical calculations of this cross
section in [2]-[4]. It has been obtained in all these papers that the
cross section is simply proportional to the pion wave function
squared: $d\sigma/dy_1\sim |\phi_{\pi}(y_1)|^2$. In such a case, it
would be sufficient to measure only the gross features of $d\sigma/dy$
to reveal the main characteristic properties of $\phi_{\pi}(x)$, and
to discriminate between various available models of $\phi_{\pi}(x).$

The purpose of this paper is to show that this is not the case.
The real situation is much more complicated, with
$d\sigma/dy$ depending on $\phi_{\pi}(x)$ in a highly nontrivial way.
We give below (in a short form) the results of our calculation of this
cross section.

\vspace{0.5cm}

\begin{figure}
\centering
\begin{picture}(300,180)(0,0)
    \SetColor{Black}
    \ArrowLine(10,125)(55,125)
    \Text(30,140)[]{\large $p_\pi$}
    \COval(75,125)(30,20)(0){Black}{White}
    \Text(75,125)[]{\large $\phi_\pi(x)$}
    \ArrowLine(80,95)(150,95)
    \Text(110,85)[]{\large $x_2 p_\pi$}
    \ArrowLine(80,155)(150,155)
    \Text(110,165)[]{\large $x_1 p_\pi$}
    \ArrowLine(195,95)(230,80)
    \Text(245,80)[]{\large $p_2$}
    \ArrowLine(195,150)(230,165)
    \Text(245,165)[]{\large $p_1$}
    \DashArrowLine(150,40)(150,90){5}
    \Text(140,65)[]{\large $q_1$}
    \DashArrowLine(190,90)(190,40){5}
    \Text(200,65)[]{\large $q_2$}
    \ArrowLine(110,10)(140,25)
    \Text(100,15)[]{\large $p$}
    \ArrowLine(200,25)(230,10)
    \Text(240,15)[]{\large $p'$}
    \COval(170,125)(40,40)(0){Black}{White}
    \Text(170,125)[]{\large $M$}
    \COval(170,25)(20,30)(0){Black}{White}
    \Text(170,25)[]{\large $g_\xi(u)$}
\end{picture}
\caption{Kinematics and notations}
\label{fig1}
\end{figure}
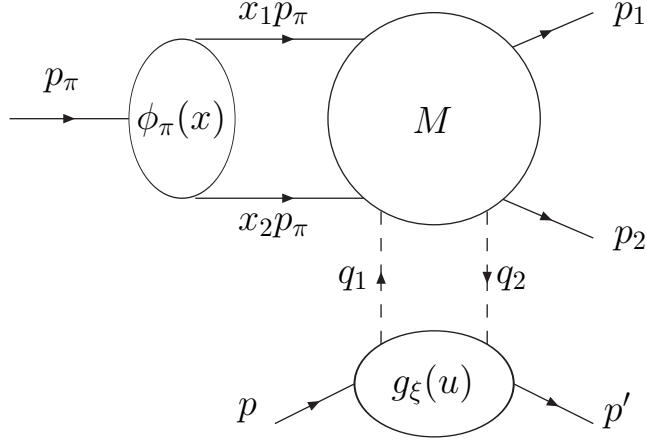

{\bf 2.} \hspace{1cm} The kinematics of the process is shown in 
fig.~\ref{fig1}.

We take the nucleon as a target, and the initial and final nucleons 
are substituted by two soft gluons with momenta $q_1$ 
and $q_2$. The lower blob in fig.1 represents the skewed gluon
distribution of the nucleon, $g_{\xi}(u)$.

The final quarks are on shell, carry the fractions $y_1$ and $y_2$ of the 
initial pion momentum, and their transverse momenta are:
 $({\bf k}_{\perp}+({\bf q}_{\perp}/2))$ and $(-{\bf k}_{\perp}+
({\bf q}_{\perp}/2)),\,  q_{\perp}\ll k_{\perp}$, where
$ q_{\perp}$ is the small final transverse momentum of the target,
while $ k_{\perp}$ is large.

The upper blob M in fig.1 represents the hard kernel of the amplitude which 
includes all hard propagators. 
For calculation of M in the leading twist approximation and in the lowest
order in $\alpha_s$, the massless pion can be substituted 
in all diagrams by two massless on shell quarks with the collinear momenta 
$x_{1}p_{\pi}$ and $x_{2}p_{\pi}$ and with zero transverse momenta, as
account of primordial virtualities and transverse momenta results only in
higher twist corrections to M. The leading twist pion wave function
$\phi_{\pi}(x,\mu_o)$ describes the distribution of these quarks in
momentum fractions $x_1$ and $x_2$.
\footnote{
As usual, on account of leading logs from loops the soft pion wave function 
$\phi_{\pi}(x,\mu_o)$ evolves to $\phi_{\pi}(x,\,\mu)\sim \int^{\mu}d^2 
l_{\perp}\Psi_{\pi}(x,\,l_{\perp})$, where $\mu$ is the characteristic scale 
of the process. And the same for the gluon distribution: $g_{\xi}(u,\mu_o)
\ra g_{\xi}(u,\mu)$.
} 

\begin{figure}
\centering
\begin{picture}(270,200)(0,0)
    \SetColor{Black}
    \ArrowLine(10,125)(55,125)
    \Text(30,140)[]{\large $p_\pi$}
    \COval(75,125)(30,20)(0){Black}{White}
    \Text(75,125)[]{\large $\phi_\pi(x)$}
    \ArrowLine(80,95)(180,95)
    \Vertex(180,95){2}
    \ArrowLine(80,155)(180,155)
    \Text(130,165)[]{\large $x_1 p_\pi$}
    \Vertex(180,155){2}
    \DashLine(180,95)(180,155){5}
    \ArrowLine(180,95)(210,80)
    \Text(225,75)[]{\large $p_2$}
    \ArrowLine(180,155)(210,170)
    \Text(225,175)[]{\large $p_1$}
    \DashArrowLine(110,40)(110,95){5}
    \Text(100,65)[]{\large $q_1$}
    \Vertex(110,95){2}
    \DashArrowLine(150,95)(150,40){5}
    \Text(160,65)[]{\large $q_2$}
    \Vertex(150,95){2}
    \ArrowLine(70,10)(110,25)
    \Text(60,15)[]{\large $p$}
    \ArrowLine(150,25)(190,10)
    \Text(200,15)[]{\large $p'$}
    \COval(130,25)(20,30)(0){Black}{White}
    \Text(130,25)[]{\large $g_\xi(u)$}
\end{picture}
\caption{One of the diagrams}
\label{fig2}
\end{figure}
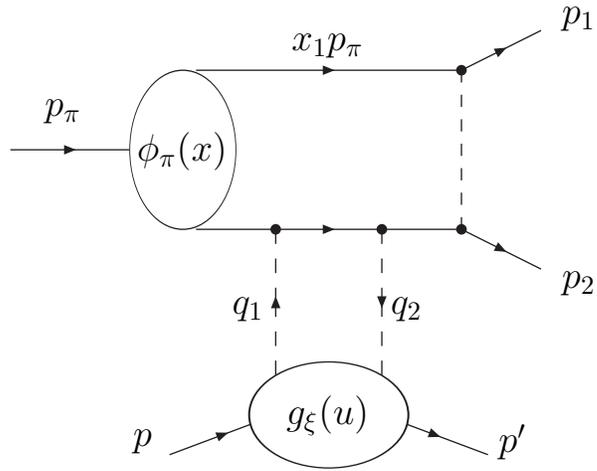

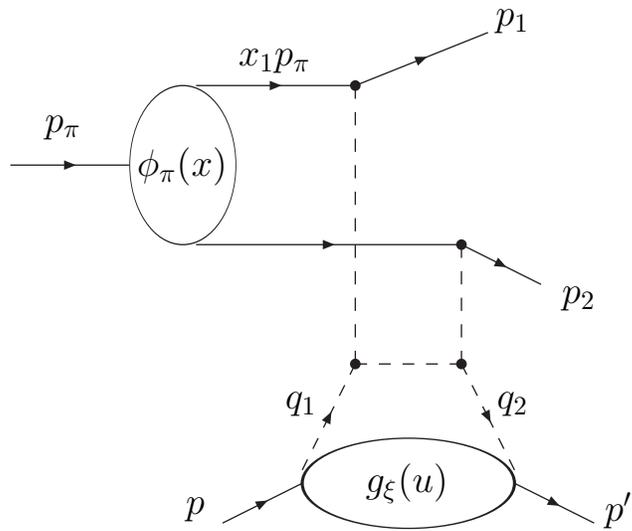
\begin{figure}
\centering
\begin{picture}(270,200)(0,0)
    \SetColor{Black}
    \ArrowLine(10,155)(55,155)
    \Text(30,170)[]{\large $p_\pi$}
    \COval(75,155)(30,20)(0){Black}{White}
    \Text(75,155)[]{\large $\phi_\pi(x)$}
    \ArrowLine(80,125)(180,125)
    \Vertex(180,125){2}
    \ArrowLine(80,185)(140,185)
    \Text(110,195)[]{\large $x_1 p_\pi$}
    \Vertex(140,185){2}
    \ArrowLine(180,125)(210,110)
    \Text(225,105)[]{\large $p_2$}
    \ArrowLine(140,185)(190,205)
    \Text(200,210)[]{\large $p_1$}
    \DashLine(140,80)(140,185){5}
    \DashLine(180,80)(180,125){5}
    \Vertex(140,80){2}
    \Vertex(180,80){2}
    \DashLine(140,80)(180,80){5}
    \DashArrowLine(120,40)(140,80){5}
    \Text(120,65)[]{\large $q_1$}
    \DashArrowLine(180,80)(200,40){5}
    \Text(200,65)[]{\large $q_2$}
    \ArrowLine(90,20)(120,35)
    \Text(80,25)[]{\large $p$}
    \ArrowLine(200,35)(230,20)
    \Text(240,25)[]{\large $p'$}
    \COval(160,35)(17,40)(0){Black}{White}
    \Text(160,35)[]{\large $g_\xi(u)$}
   \end{picture}
\caption{The diagram giving the enhanced contribution}
\label{fig3}
\end{figure}

The hard kernel M is proportional to the scattering 
amplitude of two initial collinear and on shell quarks of the pion on the 
on shell gluon: 
$$
d(x_{1}p_{\pi})+{\bar u}(x_{2}p_{\pi})+g(q_{1})\rightarrow d(p_1)+
{\bar u}(p_2)+g(q_2)
$$
In lowest order in $\alpha_s(k_{\perp})\,\,\,$  M consists of 31 
connected Born diagrams, each one is $\sim O(\alpha_{s}^{2}(k_{\perp}))$
and contains exactly three hard propagators (except for one diagram with
the 4-gluon vertex, which has only two). Two 
diagrams are represented explicitly in fig.~\ref{fig2} and fig.~\ref{fig3}. 

In the c.m.s. and to the leading twist accuracy,
the initial and final soft gluons can be considered to be on 
shell, with transverse polarizations, carrying fractions 
$(u+\xi)$ and $(u-\xi)$ of the mean nucleon momentum $\ov P$. 
\footnote{
 The small skewedness, $\xi \ll 1$, is always implied. It is typically:
$\xi \sim 10^{-2}$, in the Fermilab experiment.}
The nucleon can be considered as being massless, spinless,  
and its skewed gluon distribution
$g_{\xi}(u,t,\mu_o)=g_{\xi}(-u,t,\mu_o),\,  -1<u<1,\,$ is defined as
\footnote{
Here and below the gluon string is always implied in gauge invariant
definitions of bilocal operators, both for the gluon distribution of the
nucleon $g_{\xi}(u)$, and for the pion wave function $\phi_{\pi}(x)$.
}
:
\bq
\langle P^{\prime}|A_{\lambda}^{a,\perp}(v_1)\,A_{\nu}^{b,\perp}(v_2)|P
\rangle _{\mu_o} = -g_{\lambda\nu}^{\perp}\frac{\delta^{a b}}{8}
\int_{-1}^{1}\frac{du}{2}\,\frac{g_{\xi}
(u,t,\mu_o)}{(u+\xi-i\epsilon)(u-\xi+i\epsilon)}\times \nonumber
\eq
\bq
\times \frac{1}{2}\left ( e^{-i(u+\xi)({\ov P}v_1)+i(u-\xi)({\ov P}v_2)}+
e^{-i(u+\xi)({\ov P}v_2)+i(u-\xi)({\ov P}v_1)}\right )\,.
\eq

This is equivalent to the standard definition 
(see [5-7]): 
\bq
\langle P^{\prime}|G_{\mu\lambda}^{a}(v_1)\,G_{\lambda\rho}^{a}(v_2)|P 
\rangle _{\mu_o} =2\,{\ov P}_\mu {\ov P}_{\rho}\int_{-1}^{1}\,
\frac{du}{2}\, g_{\xi}(u,t,\mu_o)\times \nonumber
\eq
\bq
\times \frac{1}{2}\left ( e^{-i(u+\xi)({\ov P}v_1)+i(u-\xi)({\ov P}v_2)}+
 e^{-i(u+\xi)({\ov P}v_2)+i(u-\xi)({\ov P}v_1)}\right )\,.
\eq

The kinematical variables are defined as:
\bq
q_1=(u+\xi)\,{\ov P},\quad q_2=(u-\xi)\,{\ov P},\quad{\ov P}=(P+P^
{\prime})/2\,,\nonumber
\eq
\bq
\Delta=(q_1-q_2)=2\,\xi\, {\ov P},\quad \xi=\frac{\p}{2 y_1 y_2 s},
\quad z_1=\frac{u+\xi}{2\,\xi},\quad 
z_2=\frac{u-\xi}{2\,\xi}\,,
\eq
\bq 
z_1-z_2=1,\quad 2\,(p_{\pi}\Delta)=M^2=\frac{\p}{y_1 y_2}\,. \nonumber 
\eq
\vspace{0.4cm}

According to the well developed approach to  description of 
hard exclusive processes in QCD [8]-[11] (see [12] for a review), all 
hard gluon and quark lines in all diagrams (see e.g. figs.2 
and 3) have to be written down explicitly and substituted 
by their perturbative propagators. In other words, the hard momentum flow 
have to be made completely explicit and these hard lines of diagrams 
constitute the hard kernel M. They should not be hidden  
as (a derivatives of) "the tails" of the unintegrated pion wave function 
$\Psi_{\pi}(x,\,l_{\perp})$, or of the "unintegrated gluon distribution". 
This is, first of all, what differs our approach from previous 
calculations of this process in [2-4] where, besides, most of diagrams were 
either ignored or calculated erroneously. 
\footnote{
While a small number out the whole set of 31 diagrams can possibly be 
reinterpreted through "the tails" of the unintegrated pion wave function 
$\Psi_{\pi}(x,l_{\perp})$ or of the "unintegrated gluon distribution", 
this is definitely not the case for most of diagrams. 
So, we see no much meaning to proceed in this way.

In a sense, the contribution of each diagram is essential in obtaining the 
numerical description for the "y"-distribution of jets in sect.4. 
In our approach all 31 diagrams are treated on equal footing.
}

Due to this the function $\Psi_{\pi} (x,\,l_{\perp})$ ( and the 
unintegrated gluon distribution) never appears explicitly in our 
calculations. It always enters the answer for each of 31 Born diagrams 
implicitly only through the function $\phi_{\pi}(x,\mu)$, i.e. only
in the standard integrated form: $\phi_{\pi}
(x,\,\mu)\sim \int^{\mu} d^2 l_{\perp}\Psi_{\pi} (x,\,l_{\perp})$. 
\footnote{
In \cite{FMS} the authors tried to use the evolution equation for the pion
wave function to obtain its "tail". For instance, for the asymptotic wave
function it was obtained in this way: 
$\Psi_{\pi}^{asy}(x,\,k_{\perp})\sim \phi_{\pi}^{asy}(x,\,\mu)/k_{\perp}^2$. 
It remains unclear for us how it is possible to obtain such
a result from the evolution equation for the asymptotic wave function which
looks as: $d\phi_{\pi}^{asy}(x,\,\mu)/d\ln \mu=\int dy\,V(x,y)\phi_{\pi}^
{asy}(y,\mu)= 0.$}

So, the structure of the amplitude is (symbolically):
\bq
T\sim \langle P^{\prime}|A^{\perp}\cdot A^{\perp}|P \rangle \otimes({\bar 
\psi}_1 M \psi{_2}) \otimes \langle 0|{\bar u}\cdot d|\pi^{-} \rangle,
\eq
where the first matrix element introduces the skewed gluon distribution
of the nucleon $g_{\xi}(u)$, 
${\bar \psi}_1$ and $\psi{_2}$ are the free spinors of final 
quarks, "M" is the hard kernel, i.e. the product of all vertices and hard 
propagators, the last matrix element introduces the pion wave function
$\phi_{\pi}(x)$, and $\otimes$ means the appropriate convolution. 

As an example, 
let us consider the diagram in fig.~\ref{fig2}. Proceeding in the above 
described way (see the appendix), one obtains the contribution to the 
amplitude (the Feynman gauge is used for the hard gluon):
\bq
T_2= - \frac{16}{9}\,\frac{\omega_o}{y_2}\int_{0}^{1} \frac{dx_1\,
\phi_{\pi}(x)}{x_1 x_2}
\int_{-1}^{1}\frac{du\,g_{\xi}(u)}{(u-\xi)(u+\xi)}\,,
\eq
\bq 
\omega_o=\delta_{ij}\frac{(4\pi\alpha_s)^2}{96}\,f_{\pi}\,({\ov \psi}_1
{\hat \Delta}\gamma_5\psi{_2})\,\frac{(y_1 y_2)^2}{k_{\perp}^4},\quad
{\hat \Delta}=\Delta_{\mu}\gamma_{\mu},
\eq
where ${\ov \psi}_1$ and $\psi{_2}$ are the free spinors of the final quarks, 
\,$\Sigma_{spins}|{\ov \psi}_1{\hat \Delta}\gamma_5\,\psi_{2}|^2=2\,
k_{\perp}^4/(y_1 y_2)$,\, "ij" are their colour indices,
and $f_{\pi}\simeq 130\,MeV$ is the pion decay constant.

As it is expected that, for this process, the imaginary part of the amplitude
is the main one at high energy, we show explicitly in ch.3
only its value. For the diagram in fig.~\ref{fig2} this gives (in all
diagrams the terms $i\epsilon$ are introduced into denominators through
$s\ra s+i\epsilon$,\,\, i.e. $\,\xi\ra \xi-i\epsilon)$: 
\bq 
Im\,T_2=\frac{16}{9}\,\frac{2\pi\,s\,\omega_o\,y_1}{k_{\perp}^2}\,
g_{\xi}(\xi)\int_{0}^{1} \frac{dx_1\,\phi_{\pi}(x)}{x_1 x_2}\,.
\eq

As a final example, let us also consider the diagram in fig.~\ref{fig3}, as 
it gives (together with the mirror diagram obtained by $q_1\leftrightarrow 
q_2)$ the enhanced contribution 
\footnote{
The analytic form of the enhancement depends on the behaviour of the gluon
distribution $g_{\xi}(u)$, see eq.(10). With $g_{\xi}(u)\sim$ const the
enhancement is logariphmic, $\sim \ln (s/k_{\perp}^2)$, but with $g_{\xi}
(u)$ from eq.(18) it is only numerical.
}
:
\bq
T_3=\frac{\omega_o}{y_1 y_2}\int_{0}^{1}\frac{dx_1\,\phi_{\pi}(x)}{x_1 x_2 
}\int_{-1}^{1}\frac{du\,g_{\xi}(u)\,N}{(u-\xi)(u+\xi)
[z_1(x_1-y_1)-x_1 y_2]}\,\,\,,
\eq
\bq
N=[ - 8\,z_1 z_2]+(8\,y_1 y_2-3-x_1 y_1-x_2 y_2)+(z_1+z_2)(x_1-y_1)\,,
\eq
\bq
Im\,T_3=8\,\frac{\pi\,s\,\omega_o}{k_{\perp}^2}\,\int_0^1\frac{dx_1\,\phi_
{\pi}(x)\,g_{\xi}(\bar u)}{x_1 x_2|x_1-y_1|}\Theta(|x_1-y_1|> \delta)+\cdots,
\eq
\bq
{\bar u}=\xi\left ( \frac{x_1 y_2+x_2 y_1}{x_1-y_1}\right ),\quad
\delta=k_{\perp}^2/s\,,
\eq
where only the enhanced term is shown explicitly in eq.(10),
which originates from the term in square brackets in eq.(9).
\footnote{
As the $\Theta$ - function in eq.(10) excludes
the region of $x_1$ too close to $y_1$, the integral is convergent
and only some enhancement remains (see the footnote 6).

At the conditions of the E791 experiment (and even at much larger energies,
see fig.6), this enhancement is by no means sufficient to
neglect contributions of all other diagrams.
}

{\bf 3.} \hspace {1cm} Proceeding in the above described way and summing up 
the contributions of all 31 Born diagrams, one obtains for the cross section:
\bq
d\sigma_N=\frac{1}{8\,(2\pi)^5}\frac{1}{s^2}\,|T|^2\,\frac{dy_1}{y_1 y_2}
d^2 k_{\perp}d^2 q_{\perp},\quad T\simeq i\,Im\,T= 
i\,\frac{2\pi\,s\,\omega_o}{k_{\perp}^2}\,g_{\xi}(\xi)\,\Omega\,\,,
\eq
\bq
\Omega=\int_0^1 dx_1\,\phi_{\pi}(x_1)\left (\Sigma_1+\Sigma_2+\Sigma_3+
\Sigma_4\right )\,,
\eq 
\bq
\Sigma_1=\left [\,\frac{4}{x_1 x_2\,|x_1-y_1|}\,\frac{g_{\xi}({\bar u})}
{g_{\xi}(\xi)}\,\Theta(|x_1-y_1|> \delta)\,\right ]+(y_1 \leftrightarrow y_2),
\eq
\bq
\Sigma_2=\frac{1}{x_1^2\,x_2^2\,y_1\,y_2}\Biggl \{\,- (x_1 x_2+y_1 y_2)+
\Biggr. \nonumber
\eq
\bq
\Biggl. +\left [ |x_1-y_1|(x_1-y_2)^2\,\frac{g_{\xi}({\bar u})}
{g_{\xi}(\xi)}\,\Theta(|x_1-y_1|> \delta) +
(y_1\leftrightarrow y_2) \right ] \Biggr \}\,,
\eq
\bq
\Sigma_3=\frac{1}{9}( \frac{x_1 x_2+y_1 y_2}{x_1^2 x_2^2 y_1 
y_2} ) \Biggl \{ - 1+\left [  |x_1-y_1|
\frac{g_{\xi}({\bar u})}{g_{\xi}(\xi)}\,\Theta(|x_1-y_1|> \delta)
+(y_1 \leftrightarrow y_2) \right ] \Biggr \},
\eq
\bq
\Sigma_4=\,\frac{16}{9}\,\frac{1}{x_1 x_2 y_1 y_2}\,\xi\,\frac{
{dg_{\xi}(u)/du}|_{u=\xi}}{g_{\xi}(\xi)}\,.
\eq

The expressions (12)-(17) constitute the main result of this paper.
\vspace{0.3cm}

Let us note that while the separate terms in $\int dx\,\phi_{\pi}(x)\Sigma_{
2}$ are logarithmically divergent at $x_{1,2}\ra 0$, it is not difficult 
to see that the divergences cancel in the sum, so that the integral is 
finite. And the same is valid for $\Sigma_{3}$. 
This is an important point as it shows that the whole approach is
selfconsistent, i.e. the hard kernel remains hard and the soft end point 
regions $x_{1,2}\ra 0$ give only power suppressed corrections. 

{\bf 4.} \hspace{1cm} In this section we present some numerical estimates
of the cross section, based on the above expressions (12)-(17). Our  main
purpose here is to trace the distribution of jets in longitudinal momentum
fractions $y_1,y_2$ depending on the profile of the pion wave function
$\phi_{\pi}(x).$

{\bf a}) As the calculations were performed in the leading twist 
approximation which becomes applicable at sufficiently large $k_{\perp}$ 
only, we take $k_{\perp}=2\,GeV$. This  assumes that the higher 
twist effects are not of great importance at such a value of $k_{\perp}$, 
and gives a possibility of comparison with the E791-data.

{\bf b}) For the skewed gluon distribution $g_{\xi}(u,t,\mu)$ of 
the nucleon at $t\simeq - q_{\perp}^2\simeq 0$  
we use the simple form (as we need it at $|u|\geq \xi$ only, and because 
$g_{\xi}(u)\ra g_o(u)$ at $|u|\gg \xi$):
\bq
g_{\xi}(u,t=0,\,\mu \simeq k_{\perp}\simeq 2\,GeV)|_
{u\geq \xi}\simeq u^{-0.3}(1-u)^5.
\eq
This form agrees numerically reasonably well with 
the ordinary, $g_o(u,\mu\simeq 2\,GeV)$, and skewed, $g_{\xi}(u,t=0,
\mu\simeq 2\,GeV)$, gluon distributions of the nucleon calculated
in [13] and [14] respectively (in the typical region 
of the E791 experiment: $|u|\geq \xi \sim 10^{-2}$). 

The detailed consideration of nuclear effects is out the scope of this paper.
So, we simply assume that these effects result mainly in an overall factor 
(see \cite{FMS}):
\bq
d\sigma_A(t\simeq -q_{\perp}^2)\simeq d\sigma_N(t=0)\,|A\,F_A(t)|^2,
\eq
where $F_A(t)=\exp \{bt/2\},\, b=\langle R_A^2\rangle/3,$ is the nuclear 
form factor.

\begin{figure}
\centering
\includegraphics[width=\textwidth]{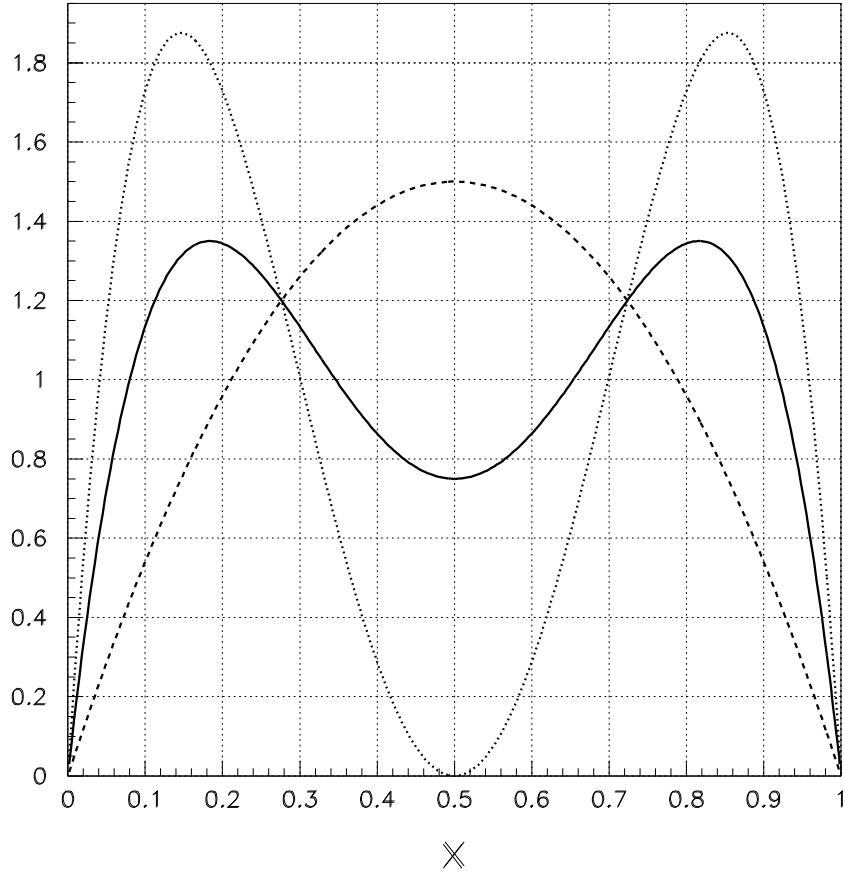}
\vspace{-7mm}
\caption{Profiles of the pion wave functions: 
a) $\phi_{\pi}^{CZ}(x,\mu \simeq 0.5\,GeV)=30\,x_1x_2(x_1-x_2)^2$ - 
dotted line;\, 
b) $\phi_{\pi}^{CZ} (x, \mu\simeq 2\,GeV)=15\,x_1x_2[0.2+(x_1-x_2)^2]$
- solid line;\, c) $\phi_
{\pi}^{asy}(x)=6\,x_1x_2$ - dashed line.}
\label{fig4}
\end{figure}

{\bf c}) As for the pion leading twist wave function, $\phi_{\pi}(x,\,\mu)$, 
we compare two model forms: the asymptotic form, $\phi_{\pi}^{asy}(x,\,
\mu)=6x_1x_2$, and the CZ-model \cite{Ch3}. The latter has the form: 
$\phi_{\pi}^{CZ}(x,\,\mu_o\simeq 0.5\,GeV)=30\,x_1x_2(x_1-x_2)^2$, at the 
low normalization point. Being evolved up to the characteristic scale 
$\mu \simeq k_{\perp}\simeq 2$\,GeV, it looks as
\footnote{
As the final quarks are free, the wave function of this two quark system is
the asymptotic one, and it does not evolve in the leading log approximation.
}
(see fig.~\ref{fig4}):
\bq
\phi_{\pi}^{CZ}(x,\,\mu\simeq 2\,GeV)=15\,x_1x_2\Bigl [(x_1-x_2)^2+0.2
\Bigr ].
\eq

\begin{figure}
\centering
\includegraphics[width=\textwidth]{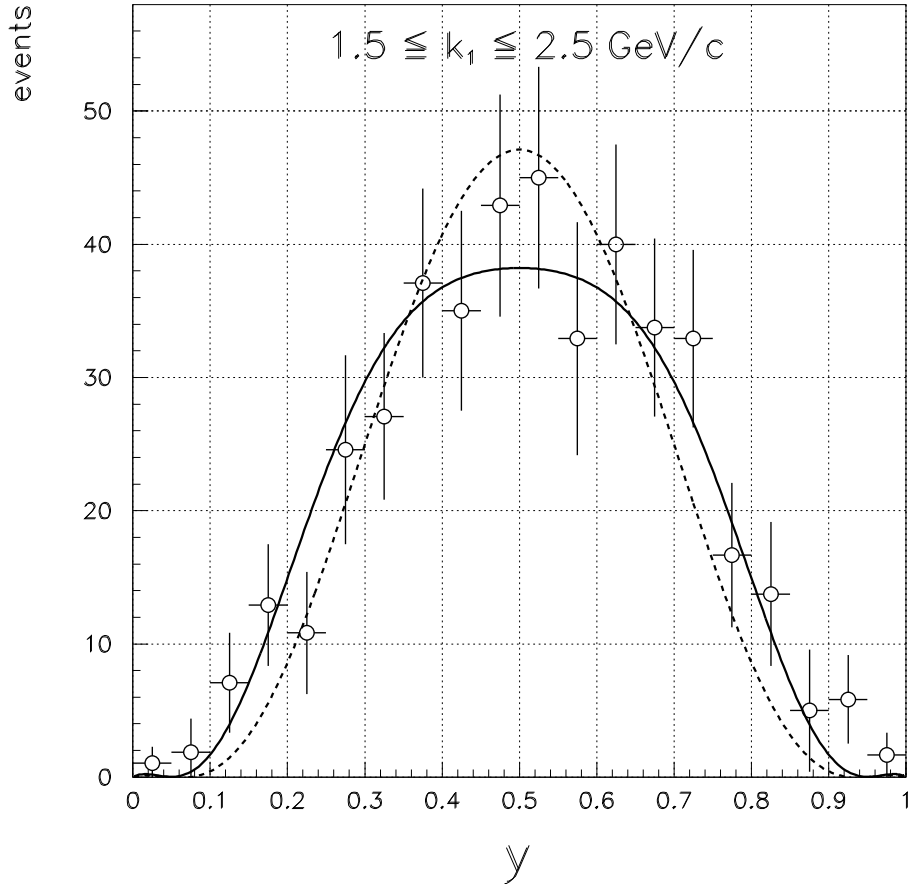}
\vspace{-7mm}
\caption{The y-distribution of jets calculated for $k_{\perp}=2\,GeV,\,
E_{\pi}=500\,GeV$ and with the pion wave functions: $\phi_{\pi}^{CZ}(x,\,
\mu\simeq 2\,GeV)$ - solid line,\, $\phi_{\pi}^{asy}(x)$ - dashed line. The
overall normalization is arbitrary, but the relative normalization of 
two curves is as calculated. The data points are from the E791
experiment~[1].}
\label{fig5}
\end{figure}

The results of these numerical calculations are then compared with the 
E791-data, see fig.~\ref{fig5}.

It is seen that, unfortunately, while two pion wave functions are 
quite different, the resulting distributions of jets in longitudinal momenta 
are similar and, it seems, the present experimental accuracy is insufficient 
to distinguish clearly between them. Moreover, even the ratio of the 
differential cross sections is not much different from unity: 
$d\sigma^{asy}/d\sigma^{CZ} \simeq 1.2$ at $y_1=0.5$, and the same 
ratio is $\simeq 0.7$ at $y_1=0.25$.

In such an unhappy situation, the theoretical calculations should be also  
performed with a maximal possible accuracy (to account additionally for:
the quark distributions, higher twist corrections, hard loop 
corrections, nuclear effects etc.). 
\begin{figure}
\centering
\includegraphics[width=.75\textwidth]{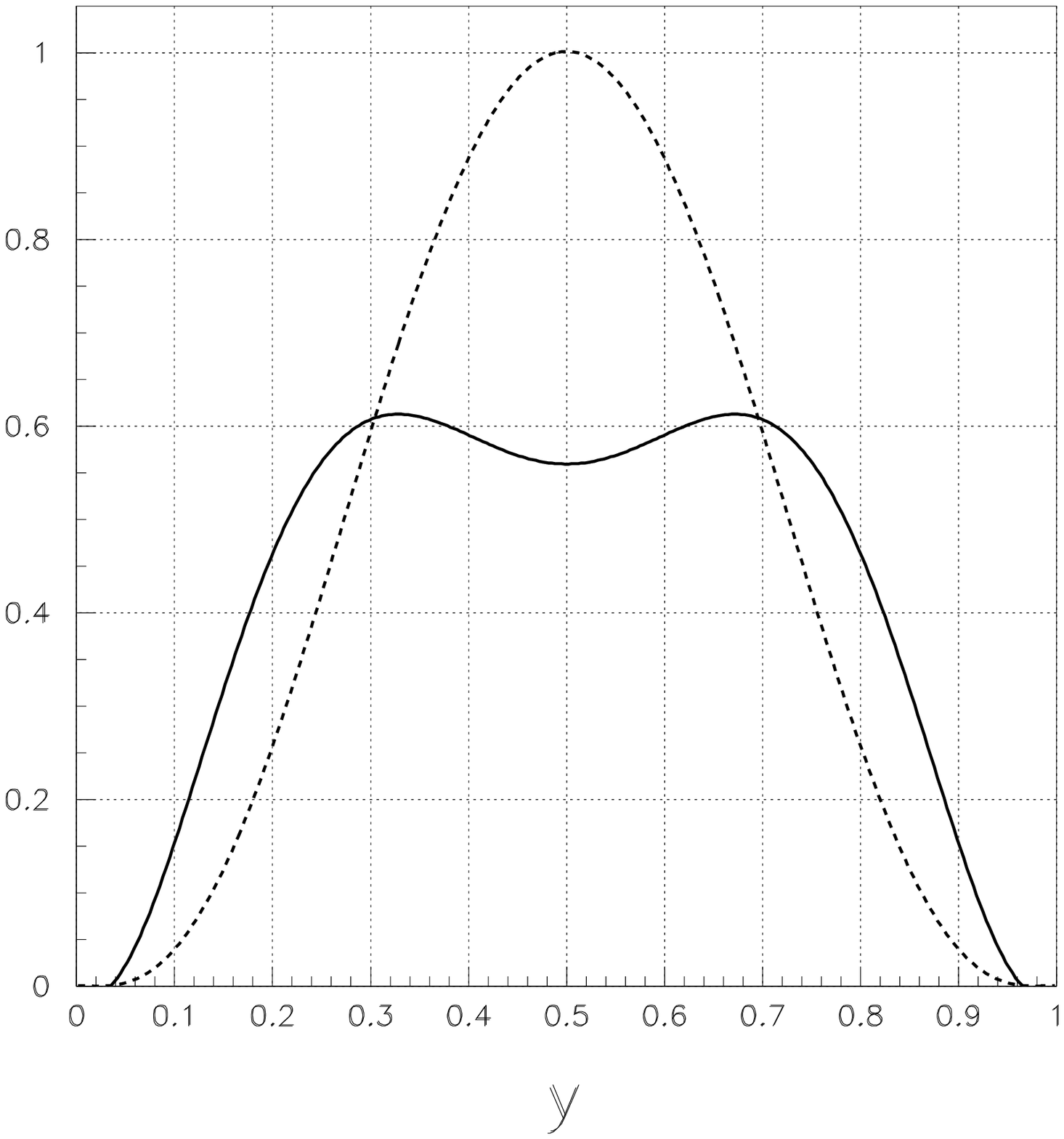}
\vspace{-5mm}
\caption{ The same as in Fig.5, but with $E_{\pi}=5\,TeV$. }
\label{fig6}
\end{figure}

We also show in fig.~\ref{fig6} 
the same distributions with the pion energy ten times larger, $E_{\pi}
=5\,TeV,\,k_{\perp}=2\,GeV$. It is seen that even this does not help much 
(as the form of the distribution weakly depends on the pion energy). 
The same ratios of the cross sections are here $\simeq 1.7$ 
and $\simeq 0.7$ respectively.

Recently the Coulomb contribution to the cross section has been calculated
in \cite{Ivan}. 
\footnote{
Since the electromagnetic contribution is real while the strong one is
mainly imaginary, they do not interfere.}
Its value for Pt is: $d\sigma_{Pt}^{electr}/dk_{\perp
}^2 dy < 10^{-5}\, mbarn\cdot GeV^{-2}$, for $E_{\pi}=500\,GeV,
\,k_{\perp}=2\,GeV$. Using the above given formulae, one obtains for the
strong cross section at the middle point $y=0.5$: $d\sigma_{Pt}^{CZ}/dk_
{\perp}^2 dy \simeq 2.5\cdot 10^{-2}\,mbarn\cdot GeV^{-2}$ with the 
same parameters, and $\simeq 0.25\,mbarn\cdot GeV^{-2}$ at $E_{\pi}=
5\,TeV$. It is seen that the electromagnetic contribution is small.
\vspace{0.5cm}

{\bf Note added:}\hspace{1cm} After this work has been completed, 
the paper \cite{Braun} on the same subject appeared. Comparison shows that,
although the qualitative conclusions are similar, the analytic expressions
for the scattering amplitude differ essentially in this paper and 
in \cite{Braun}.
\footnote{
In comparison with their original calculation the authors of \cite{Braun} 
have found recently an additional missed contribution, so that their revised 
analytic results for the amplitude agree now with those obtained 
in this paper, see eqs.(12-17).
}

\begin{center}\bf Acknowledgements \end{center}

I am grateful to V.S. Fadin for useful discussions and critical remarks. I
also thank  A.E. Bondar and B.I. Khazin for explaining me some details of
the E791 experiment.

\vspace{1cm}

\hspace{2cm}{\bf Appendix}
\vspace{0.5cm}

The purpose of this appendix is to give (somewhat schematically) some
details of how the diagrams have been calculated. As an example, let us 
consider the simplest diagram in fig.2. The momentum of the virtual gluon is:
$k=(x_1 p_{\pi}-p_1)$. The momenta of  virtual quarks
are: $\sigma=(-x_2 p_{\pi}-z_1 \Delta)$ and $\rho=(-x_2 p_{\pi}-\Delta)$,
with $q_{1,2}=z_{1,2}\Delta$,\, see eq.(3). The overall
denominator is therefore: $D=k^2\sigma^2\rho^2=(-x_1 x_2^2 y_2 z_1)(2p_{\pi}
\Delta)^3$. This diagram looks (in the operator form) as:
\bq
T_2=i\frac{(4\pi\alpha_s)^2}{D}\Biggl [{\ov d}\frac{\lambda^a}{2}\gamma_
{\mu}d\Biggr ]\,
\Biggl [{\ov u}\frac{\lambda^b}{2}\gamma_{\lambda}\,{\hat \sigma}
\frac{\lambda^c}{2}\gamma_{\nu}\,{\hat \rho}\frac{\lambda^a}{2}\gamma_{\mu}
u\Biggr ] \Biggl [A^{\perp,b}_{\lambda}\,A^{\perp,c}_{\nu}\Biggr ]\,.
\eq
The matrix element of this operator between the initial and final states is
factorized then into three parts. The first part is the matrix element of the
gluon fields between the initial and final nucleons, this introduces the
gluon distribution, see eq.(1):
\bq
\langle P^{\prime}|A_{\lambda}^{\perp, b}\,A_{\nu}^{\perp, c}|P
\rangle \ra -\, g_{\lambda\nu}^{\perp}\frac{\delta^{bc}}{8}
\,\frac{g_{\xi}(u)}{4\,(u-\xi)(u+\xi)}.
\eq
 
The second part is the matrix element
of two quark fields $d^k_{\alpha}\,{\ov u}^l_\beta$ between the pion and
vacuum, this introduces the pion wave function
\footnote{
Eq.(23) is a shorthand for a standard definition of the pion wave function
$\phi_{\pi}(x)$ through the matrix element of the gauge invariant bilocal 
quark operator, see [8,\,9], in the same way
as eq.(22) is a shorthand for eq.(1). See also the footnote 3.
}
:
\bq
\langle 0|d^k_{\alpha}\,{\ov u}^l_\beta|\pi^-\rangle \ra \frac{\delta^{k l}}
{3}\,\frac{({\hat p}_{\pi}\gamma_5)_{\alpha\beta}}{4}\,i\, f_{\pi}\,
\phi_{\pi}(x)\,.
\eq
The last part is the matrix element of two remaining quark fields between
the vacuum and the final state of two free quarks. This introduces the
Dirac spinors ${\ov \psi}_1$ and $\psi_2$. The result is integrated then
over "x" and "u". So, one obtains:
\bq
T_2=\int^1_0 dx_1\int^1_{-1}du \,(4\pi\alpha_s)^2 \frac{f_{\pi}
\phi_{\pi}(x)}{12}\frac{g_{\xi}(u)}{32\,(u-\xi)(u+\xi)}\left (\frac{4}{3}
\right )^2\delta_{ij}\left ({\ov \psi}_1 M \psi_2\right ),
\eq
\bq
({\ov \psi}_1 M \psi_2)=\frac{1}{D}\left ({\ov \psi}_1 \gamma_{\mu}
{\hat p}
_{\pi}\gamma_5\gamma^{\perp}_{\lambda}{\hat \sigma}\gamma^{\perp}_{\lambda}
{\hat \rho}\gamma_{\mu}\psi_2\right )=4\, x_{2} z_{1} \frac{(2 p_{\pi}\Delta)}
{D}\left ({\ov \psi}_{1}{\hat \Delta}\gamma_5 \psi_{2}\right )\,.
\eq

On the whole, one obtains the eqs.(5,6). All other 30 diagrams have been 
calculated in the same way.


\begin{thebibliography}{25}
\bibitem{Aitala}
E.M. Aitala et. al. (E791 Collaboration), hep-ex/{\bf 0010043}\\
D. Ashery, hep-ex/{\bf 9910024};  Invited Talk at X Intern. Light-Cone \\
Meeting, Heidelberg, June 2000: hep-ex/{\bf 0008036}
\bibitem{Br}
G. Bertsch, S.J. Brodsky, A.S. Goldhaber and J.G. Gunion,\\
Phys. Rev. Lett. {\bf  47} (1981) 297
\bibitem{FMS}
L. Frankfurt, G.A. Miller and M. Strikman, Phys. Lett. {\bf B304} 
(1993) 1; Found. of Phys. {\bf 30} (2000) 533 (hep-ph/{\bf 9907214});
hep-ph/{\bf 0010297}
\bibitem{Nik}
N.N. Nikolaev, W. Schafer and G. Schwiete, Phys. Rev. {\bf D63} (2001) 
014020; hep-ph/{\bf 009038}
\bibitem{Rob}
D. Muller, D. Robaschik, B. Geyer, F.-M. Dittes and J. Horejsi,\\
 Forts. Phys. {\bf 42} (1994) 101
\bibitem{Rad}
A.V. Radyushkin, Phys. Lett. {\bf B385} (1996) 333; Phys. Rev. {\bf D56}
(1997) 5524
\bibitem{Ji}
X. Ji, Phys. Rev. Lett. {\bf 78} (1997) 610; J. Phys. {\bf G24} (1998) 1181
\bibitem{Ch1}
V.L. Chernyak and A.R. Zhitnitsky, JETP Lett. {\bf 25} (1977) 510; Sov. J.
Nucl. Phys. {\bf 31} (1980) 544
\bibitem{Ch2}
V.L. Chernyak, V.G. Serbo and A.R. Zhitnitsky, JETP Lett. {\bf 26} (1977)
594; Sov. J. Nucl. Phys. {\bf 31} (1980) 552
\bibitem{Efrem}
A.V. Efremov and A.V. Radyushkin, Phys. Lett. {\bf B94} (1980) 245;\\
Teor. Math. Phys. {\bf 42} (1980) 97
\bibitem{LB}
G.P. Lepage and S.J. Brodsky, Phys. Lett. {\bf B87} (1979) 359; Phys. Rev.
{\bf D22} (1980) 2157
\bibitem{PR}
V.L. Chernyak and A.R. Zhitnitsky, Phys. Rep. {\bf 112} (1984) 173
\bibitem{GRV}
M. Gluck, E. Reya and A. Vogt, Eur. Phys. J. {\bf C5} (1998) 461
\bibitem{Rys}
K.J. Golec-Biernat, A.D. Martin and M.G. Ryskin, \\ Phys. Lett. {\bf B456}
(1999) 232;  hep-ph/{\bf 9903327}
\bibitem{Ch3}
V.L. Chernyak and A.R. Zhitnitsky, Nucl. Phys. {\bf B201} (1982) 492
\bibitem{Ivan}
D.Yu. Ivanov and L. Szymanowski, hep-ph/{\bf 0103184} 
\bibitem{Braun}
V.M. Braun, D.Yu. Ivanov, A. Schafer and L. Szymanowski, \\
hep-ph/{\bf 0103275}; Phys. Lett. {\bf B509} (2001) 43
\end{thebibliography}
\end{document}